# Connecting research in physics education, curriculum decisions and teaching practices

**Symposium Organized by GTG Physics Education Research at University-PERU**
**J Guisasola[1], J Ametller[2], D Baccino[3], M Carli[4], S Kapon[5] , A C Marti[6] , M Monteiro[7] , O Pantano[4], T Peer[5,8], P Sarriugarte[1], M Schvartzer[5] , A Suárez[3] and K Zuza[1]**

[1]Department of Applied Physics, University of the Basque Country (UPV-EHU) and Donostia Physics Education Research Group (DoPER), Spain.
[2] Department of Specific Didactics, University of Girona, 17004 Girona, Spain
[3] Consejo de Formación en Educación, ANEP, Montevideo, Uruguay
[4] Department of Physics and Astronomy, University of Padova, Padova, Italy
[5] Faculty of Education in Science and Technology, Technion – Israel Institute of Technology, Israel
[6] Instituto de Física, Universidad de la República, Montevideo, Uruguay
[7] Facultad de Ingeniería, Universidad ORT Uruguay, Uruguay
[8] Acheret Center - Multicultural Research Fellowship, Kibbutz Cabri, Israel

**Abstract.** In the symposium contributions we discuss research in physics education and the consequences of its results for physics teaching. The symposium presents four different aspects of physics teaching and learning, but all of them have research-based problem analysis in common. The problems analysed cover different aspects of the physics teaching-learning process. Innovative aspects such as the effect on learning of the integration of engineering projects in the science teaching process, the influence on the learning process of conceptions about science and attitudes, and aspects related to teaching contents and students' learning difficulties. Its conclusions are not merely intuitive proposals based on teaching experience, but on a careful planning of data collection, analysis of results and empirical basis.

## 1. Introduction

Regarding research results in any field, their transfer into practice is not necessarily straightforward. In the Girep Thematic Group PERU we work in an international cooperation effort to transfer the results of research in physics education to teachers and curriculum designers (Guisasola 2019). One of the principal objectives of the Girep Thematic Group PERU is to share Physics Education Research results and its relevance to physics education and classroom practice. In particular, to suggest ways in which the facilities for the study of physics at introductory physics courses might be improved.

Since 2012 PERU thematic group has organized symposia in Conferences and Seminars with the general objective of bringing together significant experiences and points of view from different areas of the world that are expressed in simple language, with the aim also of encouraging the application of innovative classroom practices and the implementation of research-based teaching initiatives. This chapter presents four different aspects of physics teaching and learning, but which have in common research-based problem analysis. The results of the research suggest the presence of very different factors that influence the teaching of physics and that makes this task complex. This rejects a simplistic conception of physics education that considers it a simple task that would consist of mastering the contents and having 'diplomacy' to deal with the students. On the contrary, as we will see in the symposium, the results accepted by the international community of physics teachers indicate that the task to be developed and the problems to be faced are sufficiently complex to constitute a field of research with multiple dimensions.

The four problems analysed offer a broad overview of the lines of research in PER such as students' attitudes towards the study of physics, students' epistemological beliefs and difficulties in the use of mathematical concepts, formal reasoning and conceptual learning. The common aim of the symposium is to show that the different lines of research in PER are complementary and that they analyse the different aspects that need to be taken into account in the complex task of teaching physics.

In the study developed by Kapon et al. (section 2) students' interest in learning physics in a STEM context is analysed. The research shows the possibilities and, the challenges of involving students in engineering projects as a place for learning physics at the advanced high school level are discussed. The study developed by Arturo C. Marti and colleagues (section 3), compares the attitudes and beliefs about science at the beginning of their university careers of two groups of students: physical sciences and life sciences. Some of the possible causes of the differences found and their implications for teaching are discussed. The research carried out by Ornella Pantano et al. (section 4) and P. Sarriugarte et al. (section 5) is in the more traditional line of PER research on student learning difficulties. Pantano et al. compare students' ability to answer questions about derivatives, integrals and vectors in a purely mathematical context and in the context of physics. The usefulness of the tool for both students and teachers is discussed. Sarriugarte et al. show the difficulties encountered by first-year university students in understanding the moment of inertia in the phenomenon of rotation of a rigid body around a fixed axis.

## 2. Learning Physics while Engaging in an Engineering Project

Several studies have suggested that the integration of engineering projects into the instruction of science can enhance students' attendance and engagement and support the learning of the traditional content of science. However, some studies have indicated that the learning of scientific content and practices through engagement in engineering projects is not straightforward and requires the additional design of various scaffolds. For example, some studies have reported students' tendency to engage in fulfilling the goals of their engineering design challenges, and only inconsistently engaging with the related math and science content (Berland & Steingut 2016). Others have cautioned that such activities can turn into arts and crafts activities, in which students and teachers focus on getting to a working solution by trial and error that is disconnected from the targeted science.

The problem described above is considered in the literature to reflect the "mixed success" of integrated STEM education (Berland & Steingut 2016). Berland and Steingut argued that students' perception of the value of math and science content for engineering predicts their efforts to integrate math and science content into their engineering coursework. The instructional implications they derived from this conclusion were that teachers and curriculum designers should mediate and emphasize the value of math and science to meeting these engineering challenges (Berland & Steingut 2016).

While we agree that 'value' is an influencing factor, in our view the reasons for the "mixed success" go beyond students' sense of 'value'. We suggest a complementary explanation situated in the very essence of the activity the students engage in. We will present findings from a recently published study (Kapon et al. 2021) that illustrate how learning physics by engaging in an engineering challenge generates tensions related to fundamental incongruencies between doing engineering and doing physics, and how these tensions affect students' learning. This explanation entails somewhat different instructional implications, which will be discussed in the concluding section.

### 2.1. Conceptual framework and goals

We examine the nature of learning physics through engagement in an engineering-maker project as a case of participating in a particular figured world. Figured worlds are "socially and culturally constructed realms of interpretation, in which particular characters and actors are recognized, significance is assigned to certain acts, and particular outcomes are valued over others. Each is a simplified world populated by a set of agents /…/ who engage in a limited range of meaningful acts or changes of state /…/ as moved by a specific set of forces." (Holland et al. 1998). We argue that the figured world of Engineering Maker-Based Inquiry (EMBI) in physics differs from the figured world of authentic scientific inquiry in physics and the figured world of traditional learning of physics in school, and that these differences have implications that affect students' learning of physics.

### 2.2. The study

The data were drawn from an extended ethnographic case study (18 months, 3 hours a week) of the authentic working sessions of two 11th grade students who were working on a long term engineering challenge as part of their mandatory learning of physics at the advanced high school level (Kapon et al. 2021). The physics teachers in the school in which the study took place are part of the Acheret Center (ACHERET 2005). Acheret is a community of physics educators that aims to incorporate authentic physics inquiry in schools. The project represented 40% of the requirements for matriculation in advanced level physics. The two students who worked together on the project were very different in terms of their academic achievement. The male was a low achieving student and the female was a high achieving, ambitious student. The students worked on the project in the school makerspace, and were mentored by a physics teacher (3rd author) who was trained in Acheret to serve as a research mentor in this context. The students and the mentor decided together that the goals of the project would be to design and build a working model of a solar panel that tracks and follows the movement of the sun and evaluate its performance by comparing the power efficiency of a stationary and tracking solar panel. Figure 1 presents an overview of the project.

The participants in the study were the two students and the educational staff: the teacher who mentored the students' inquiry project (a senior member in the Acheret community), the maker-space coordinator who assisted the teacher and served as the laboratory technician, the physics department head of the school (a senior member of Acheret) who helped the students to wrap up the project and write the final report in the second year when their mentor was on sabbatical, and the students' "regular" classroom physics teacher (a junior member of Acheret), who also stepped in to help during the second year. The data included videotaped observations of the authentic working sessions every second or third week over a full school year, including the oral exam in the second year, interviews with the students and the educational staff, and the artefacts that were generated during the inquiry (notes, final research report, etc.).

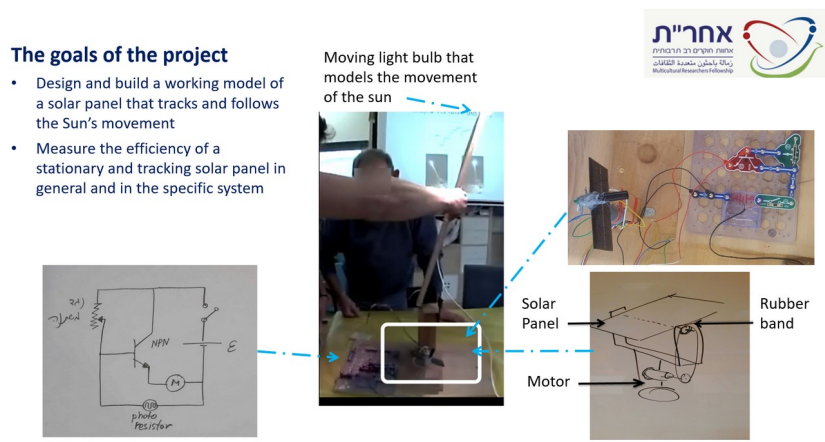

**Figure 1.** Learning Physics while Engaging in an Engineering Project

The analysis combined ethnographic accounts and fine-grained socio-linguistic discourse analysis of selected episodes. The discourse analysis of mentor-mentee interactions in the authentic working sessions was complemented by interviews and other ethnographic accounts. In a full paper that summarizes the analysis of this case study [8] we (1) characterized the figured world of EMBI, (2) identified the central legitimate forms of participation that were enacted by the students and influenced their learning, (3) articulated how these forms of participation were socially communicated, constructed, and enforced over time in the interaction between the two students and the educational staff, and (4) examined how these forms of participation facilitated (or impaired) the learning of the content and practices of the related physics.

In this chapter we use this analysis (Kapon et al. 2021) and other case studies we have conducted in physics classrooms at the Acheret center (Kapon 2016) to compare and contrast the figured world of EMBI with the figured world of authentic inquiry in physics in schools associated with Acheret Center, and with the figured world of physics in school which also shaped the students' experience. Then we compare and contrast two legitimate forms of participation in EMBI that we identified in the case study [8] and discuss the implications for the teaching and learning of physics.

### 2.2. Analysis

Table 1 compares the goals and the meaningful acts in the three laminated figured worlds we documented in Acheret: engineering maker-based inquiry (Kapon et al. 2021), authentic physics inquiry (Kapon 2016), and traditional school physics. As Table 1 shows, while there are some overlaps between the meaningful acts of EMBI and authentic scientific inquiry, there are also many differences in the goals and meaningful actions that can generate tensions, and there is a considerable difference between these figured worlds and the figured world of mainstream school science. Based on the sociolinguistic discourse analysis of the conversations in the authentic working sessions of the EMBI project (Kapon et al. 2021), we identified two distinct legitimate forms of participation in the EMBI project. Table 2 compares and contrasts them.

**Table 1.** Laminated Figured world

|  | Engineering Maker-Based Inquiry in physics (Acheret) | Authentic scientific inquiry in physics (Acheret version) | Learning physics in school |
|---|---|---|---|
| Goal | Design and build a working engineering device, evaluate its performance and improve it accordingly. | Produce reliable knowledge and explanations about the physical world. | Pass the external matriculation exam in advanced level physics, for which the project counts 40%. |
| Meaningful acts | • Characterize what the device should do and translate these characteristics into scientifically measurable features<br>• Design the device by employing scientific principles as well as | • Generate theory and quantitative models that can explain and produce testable hypotheses, and iteratively improve the models.<br>• Design experimental set up, build and | • Attend working sessions in the laboratory<br>• Submit a research proposal and final research report for external evaluation<br>• Defend the work in an externally |

| | | |
|---|---|---|
| incorporating practical considerations such as the availability of materials, how easy it is to work with them, etc.. <br> • Build a working porotype, engaging in its construction and troubleshooting. Design and perform experiments that test the performance of the device and improve the design if necessary. | improve it <br> • Learn to use and choose wisely between a variety of measurement and data processing tools, by applying considerations such as uncertainty, direct vs. indirect measures, etc. <br> • Deal with ambiguity and gaps between the experimental results and theory. Present the findings in a way that peers and experts can understand and evaluate them. | evaluated oral exam <br> Present conceptual proficiency and understanding |

While both forms of participation contributed greatly to the success of the EMBI, only the participation as an engineer contributed to the students' learning of the related physics in terms of the figured world of learning physics in school, as well the figured world of authentic scientific inquiry. Our longitudinal analysis [3] shows that at the beginning of the project both students were participating as engineers as well as technicians. However, as the project progressed and developed, the high achieving student mainly participated as an engineer, while the participation of the low achieving student was limited to participation as a technician. As the year unfolded, the low achieving student gradually became an apprentice to the laboratory technician instead of to the research mentor.

**Table 2.** Legitimate forms of participation in EMBI.

| *Participate as an engineer* | *Participate as a technician* |
|---|---|
| Engage in the following actions while employing scientific as well as practical considerations: <br> • system characterization and design <br> • troubleshooting the system <br> • testing system performance (designing and conducting experiments, interpreting results) <br> • achieve a robust understanding of the underlying scientific principles and employ them in the design and evaluation of the system | The engagement is limited solely to practical and functional considerations <br> • construction <br> • repair of technical faults (disconnected wiring, mechanical jamming and the like) <br> • receive rather than generate explanations <br> • understand the scientific principles underlying the system solely at a functional level |

Our findings show that the low achieving student developed a deep sense of ownership and pride about the project. He contributed significantly to it by constructing and stabilizing the physical model of moving "sun" (see Figure 1), and contributed to the mechanical construction of the moving the solar panel. However, our data also demonstrate that he did not understand some of the very basic physics principles that underpinned the engineering design. The student referred to this dissonance in an interview after the oral exam: "He (the external examiner) did not ask about the practical side. I was really disappointed /.../ because we worked on it, and prepared the axes, and the angles, and we cut the wood. This is the part I loved. The part I enjoyed most, that was most mine. So I was little disappointed." His partner also brought it up in the last interview: "/.../ The truth is that in this project, ah, Ram was more, he was doing the basic technical work, because... Well, now, in a retrospect, this does not seem fair. I knew electricity, so we ((the student and the research mentor)) kind of did this part quickly, and left the explanations to Ram to the end. So... / It was much more fun doing it with Ram, although he did not do much in the physics part." We interpret this as an unequitable opportunity to engage in doing physics.

### 2.3. Conclusion
Engineering maker-based inquiry offers inclusive pathways into physics that can nurture an emotional connection to physics for a more diverse body of learners. However, while supporting students' sense of belonging, which is a worthy goal in itself, it does not necessarily facilitate meaningful learning of physics. To achieve this, its implementation in physics classrooms requires careful pedagogical design that engages *all* the students as engineers and makes sure that no student will participate solely as a technician.

## 3. What are the differences in the attitudes and beliefs about science of students in the physics-mathematics and life sciences areas? and what are their impact on teaching?

It has long been accepted in the physics-teaching community that, in addition to the specific contents, the set of ideas, assumptions and previous conceptions about science, in general, the epistemological beliefs, strong influence the transformation that occurs as students move through the educational system. This specific field of epistemological attitudes and beliefs does not escape the claim of researchers to obtain quantitative information. In that sense, in the last twenty-five years a series of questionnaires or tests standardized to assess the epistemological attitudes and beliefs of the students have been proposed in the literature. There are many open questions in relation to the attitudes and beliefs of students, especially those related to the comparison between different groups, the causes of possible differences and their impact on learning. Several quantitative studies have been performed in recent years on the changes that occur in students' attitudes and beliefs depending on their previous training, the types of courses, and the teaching strategies used, as well as on the relationship between the pre-test and academic performance, among other aspects. In this section we address results recently published in the specialized literature about this topic.

### 3.1. Attitudes and beliefs of physical sciences students in comparison with other groups

One aspect regarding attitudes and beliefs about science that has received considerable attention is the comparison of the performance of different student groups as a function of the academic options adopted in the early stages of their careers. In a recent study (Suarez et al. 2022), the attitudes and beliefs about science of students of physical sciences (physics and mathematics) were compared with those of life sciences (biology and biochemistry) students at the beginning of their university degrees using the CLASS toorl (Adams et al. 2006). It is worth noting, that both groups received similar physics courses during their high-school education. The differences in performance in each of the areas that make up the questionnaire were examined and, perhaps the flashiest result, is that a larger percentage of life science students (higher than that of physical science students) adopted a "novice" behavior in problem solving.

This research was carried out with university students from the School of Sciences of Universidad de la República (Montevideo, Uruguay) who were taking General Physics I in the first semester of the undergraduate courses in Physics and Mathematics and Biology and Biochemistry. It is relevant to point out that in Uruguay, primary, secondary and pre-university schools are characterized by a common curriculum framework for all the educational institutions in the country. The usual bibliography includes algebra-based "College Physics" textbooks such as those widely used worldwide (Serway, Resnick among others). Upon completing high school, students can opt for different university degrees. The focus of this work was on two sets of recently admitted students in the first year of the aforementioned School of Science. One set is comprised by those who pursue bachelor's degrees in Physics and Mathematics (hereinafter, "physical sciences" or "PhS") and the other by those who pursue bachelor's degrees in Biology and Biochemistry (hereinafter, "life sciences" or "LS").

The authors proposed to all the students to report online their degree of agreement or disagreement with the 42 statements of the CLASS test according to a Likert scale. Some examples of the statements are the following:

- "A significant problem in learning physics is being able to memorize all the information I need to know."
- "It is useful for me to do lots and lots of problems when learning physics"
- "After I study a physics topic and feel I understand it, I have difficulty solving problems on the same topic."
- "When solving a physics problem, I look for an equation that uses the variables given in the problem and I substitute the values."
- "I enjoy solving physics problems"

Of a total of 42 statements, 27 of them are grouped into 8 categories: a) Real world connection, b) Personal interest, c) Sense making / Effort, d) Conceptual comprehension, e) Applied conceptual understanding, f) Problem solving / general, g) Problem solving / confidence, h) Problem solving / sophistication. The remaining 15 questions are not categorized. There is no agreement among the experts on some of the latter, while one in particular is used to rule out inconsistent responses. A question can be categorized in more than one category. For each category, and for the set of questions where there is agreement among the experts (36 of the 42 statements), the percentage of student responses that agree with that of the experts (favorable responses) were reported and evaluated using statistical tools as the non-parametric Mann-Whitney U test, the Cohen effect size and contrasted with a null hypothesis to be rejected if the probability is smaller than a given threshold.

The differences in performance in each of the areas of the questionnaire reveals some interesting aspects. Among other findings stands out that a considerable percentage of life science students (higher than that of physical science students) adopted a novice type of behavior in problem solving. From the analysis of the responses in the different categories, the authors highlight that a significant percentage of life sciences students try to solve physics problems following the strategy of "finding the right equation and substituting", known as "plug and chug". Being aware of this type of thinking allows us, as teachers, to anticipate the problems of our students, to stand differently in the classroom, and to develop actions aimed at changing this type of thinking, which will result in a better-quality science education. In this sense, it is important to keep in mind that in every action or omission that we make in the classroom (and that is part of the hidden curriculum), we are directly or indirectly affecting the epistemological beliefs of our students.

### 3.2. Physics (and future) teachers' attitudes and beliefs about science

The epistemological conceptions of the Uruguayan Physics teachers and future Physics teachers were also analyzed through the application of the CLASS test (Suarez et al. 202021). This study was focused on identifying categories in which there are significant similarities or differences between the two groups and also in comparison with the experts' opinions taken as reference. The importance of this study is related to the fact that aspects which exhibit positive or negative variations between the opinions of future teachers and present teachers suggest that these specific training resulted favorable or unfavorable. Moreover, the areas where the divergences with the opinions of the experts are notorious indicate that they should be reinforced.

The survey was open online to physics teachers and students from all over Uruguay for a month in the first half of the (pandemic) 2020. In relation to the teachers an electronic announcement was sent to almost all active Physics teacher in the country. The responses received were 143, nearly equally distributed by genre. The age of the respondents was distributed as 31% between 31 and 40 years, 22% declare in the age groups contiguous to the previous one (18-30 and 41-50), 25% register over 51 years of age. Regarding future physics teaching, first-year students of the Physics schools were contacted directly by their teachers in several of the schools. 138 students participated, approximately two thirds women. All the responses were analyzed using the electronic spreadsheets available in the well-known site *Physport*.

Several points can be outlined from this study. The first one is that, not surprisingly, there is a significant difference in the averages of teacher and student responses (80% for teachers and 67% for students). The categories that present the most agreement with the reference opinion are the same for teachers and future teachers. These categories are Sense making/Effort and Confidence in problem solving. In contrast, the greatest observed differences can be found in the categories Applied conceptual understanding and Sophistication in problem solving. Assuming that it is possible to consider the two groups of surveyed individuals as part of a pseudo-longitudinal approach, these differences could be hypothetically attributed to the systematic training processes and the teaching-learning experiences.

Although this study provides interesting details several open questions about the attitudes and beliefs about science of Physics teachers remain open. Some of these questions could be at what stage do student teachers transform their epistemological beliefs, approaching that of experts. It would be also interesting to know how do these attitudes impact in the teacher training processes and how do they transmit this grounding to their students.

### 3.3. A comparison with other studies recently published

As mentioned before several studies about attitudes and belief about sciences were published in the literature. A meta-analysis published in 2015 summarized 24 results available until this date based on the application of the CLASS and MPEX tests in several universities, mainly from U.S.A, Canada and United Kingdom, to students who took calculus-based physics courses . Among the main points highlighted in this meta-analysis, we can mention that students with expert-like opinions tend to choose physics majors more probably than other options. Also, there is some correlation between incoming beliefs and their gains on conceptual tests. Regarding the overall punctuation obtained in the CLASS test in these international studies and those mentioned in the previous section the comparison does not suggest large differences.

Another study in relation to the attitudes and beliefs of physics teachers was conducted in Thailand, a country with very different demographic, social, economic and cultural characteristics from those of Uruguay. In the study the responses of 196 physics teachers and 211 secondary students were analyzed. Despite the differences mentioned between the contexts of both teaching groups, the answers were very similar in the majority of the questions and categories of the CLASS. Only in two of the 42 questions of the test, significant differences are found between teachers from Thailand and from Uruguay. A similar situation was found in relation to physics teacher students in these countries (Suwonjandeen et al. 2018) who also show great differences with the experts precisely in the same questions which, in turn, present the highest degree of discrepancy between the responses of physics teachers and physics teacher students.

Very recently, a study about the epistemological beliefs and attitudes about science among Malaysian students was published (Ibrahim et al. 2022). This study was motivated by the poor student's performance and it discuss the role of the epistemological beliefs in this phenomenon. With this objective a specific tool was developed adapting the MPEX and CLASS tests. These findings suggest that positive epistemological beliefs correlate with a better performance.

To conclude this section, a couple of remarks can be stated. First, that all the previous works suggest that attitudes and beliefs about sciences play a significant role in the teaching and learning of physics. Although this role is often neglected ("the hidden curriculum"), it is manifested in several stages. Secondly, the attitudes and beliefs of physics teachers strong influence those of the students. Finally, it is clear that there are many open questions about this topic that deserve to be addressed.

## 4. Test of Calculus and Vector in Mathematics and Physics: A research-based tool for improving the teaching and learning of physics in first-year courses

### 4.1 Background

Mathematics has a paramount importance in physics education, not only because it provides the language and the technical tools for describing and making predictions about phenomena, but also due to its structural role (Uhden et al. 2012). However, in many Science and Engineering courses, mathematics is seen merely as a prerequisite for physics. Courses like calculus and algebra are supposed to provide students with mathematical tools, which students are then expected to proficiently use in their physics courses in the following semester. This belief contrasts with findings in physics education research, which show quite clearly that using mathematics in physics is different from using it in a purely mathematical context. Even students who complete a calculus course successfully may have difficulties in using the same mathematical tools in a physics context (Redish and Kuo 2015).

With the aim of raising awareness on the issue and supporting students in overcoming these difficulties, in 2018 we designed an assessment tool, the *Test of Calculus and Vectors in Mathematics and Physics* (TCV-MP). Its development and validation were described in detail in a previously published paper (Carli et al 2020). In this contribution, after retracing the history of the instrument, we describe how its use has evolved at our University and beyond through the years.

### 4.2 Test design and validation

The TCV-MP was developed in the framework of a bigger initiative called "FisicaMente", aimed at supporting first-year Science and Engineering students as they approached their "Physics 1" course. The test focuses on three mathematical tools - derivatives, integrals, and vectors - which are pivotal for any university-level physics course, and with which students have recognized difficulties. These challenges were confirmed by a survey conducted in the initial phase of the project, involving physics instructors.

Given the large target population (students from 23 courses), the instrument was designed as a multiple-choice test. Physics Education Research (PER) informed the design of the items, in combination with an analysis of students' exams. From this analysis we identified the main students' difficulties related to the use of derivatives, integrals and vectors, and we related them to those documented in the literature to identify the specific subtopics and distractors for each item. We drew upon both existing concept inventories and topic-specific literature (for further details see Carli et al. 2020).

For each subtopic, different representational formats (words, graphs, formal language, and numbers) were considered, as students may adopt different problem-solving strategies depending on the representational format used in the problem (Ibrahim& Rebello 2012). Multiple versions of each item were developed, involving different combinations of the representations used in the question and answers. Finally, for each subtopic and combination of representations two parallel items were constructed, one in the context of mathematics and one in the context of physics, in order to compare students' answers in the two contexts. The chosen physics context was mechanics, as it is the common core of all Physics 1 courses. Through this process we constructed an initial pool of 78 items, which were administered to a pilot group of 70 students. Following item analysis and student interviews, 17 pairs of parallel items were selected and included in the final version of the test. The test was administered to first-year students in Science or Engineering at the beginning of the Spring semester in 2018. Participation was voluntary and physics instructors were in charge of students' involvement. In total, 1252 students completed the test.

*4.3 Results from the first administration of the test*

We analyzed students' answers both at the whole test level and for individual items. As an example of the results that we obtained, here we discuss students' answers on items 2 and 3, regarding the topic of derivatives. In item 2M (context of mathematics), students were given the graph of a function and were asked to select the correct graph displaying its derivative; in its parallel physics item (2P), a position-time graph was provided, and students had to select the corresponding velocity-time graph. In item 3M, students were tasked with calculating the first derivative of a function at a specific point based on its graph; in the parallel physics item (3P), they had to calculate an object's velocity at a given instant based on its position-time graph. In Fig. 1 we compare students' answers in items 3M, 3P and 2P.

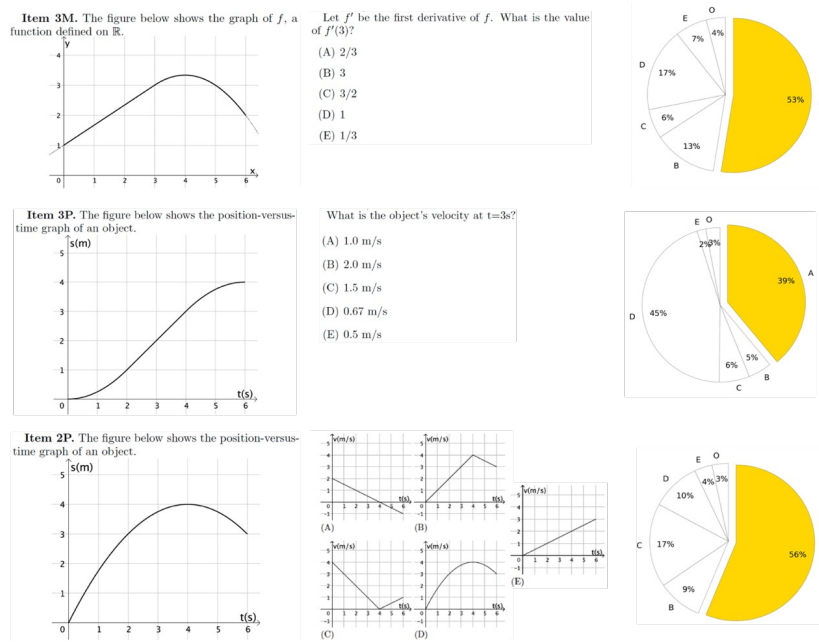

**Fig. 1.** Comparison of students' answers across two parallel mathematics/physics items (items 3M and 3P) and across two items regarding the same topic, but with different representations (items 3P and 2P).

The answer profile was quite different between items 3M and 3P, with many more students choosing the most common incorrect answer (option D, corresponding to calculating y/x or s/t) in physics than in mathematics.

This result suggests that the mathematics content of the item was only part of the problem; instead, the common incorrect conceptualization of velocity as "space over time" probably played a crucial role in students' selection of distractor D in physics. On the other hand, the comparison between items 3P and 2P offers some insights into the role of different representations. The percent of correct answers in item 2P, which did not involve calculation, was higher than in answer 3P and more aligned to that of item 3M. It was also comparable with item 2M (not shown, 64%).

### 4.4 Evolution beyond the initial administration

Our results support the need, advocated by PER, for the explicit training of students' mathematization skills in physics. As a possible means of addressing this need, in 2019 we developed online learning modules, available to students upon completing the test. Organized as "chapters" within a Geogebra book, each module corresponds to one of the topics and encompasses an introduction, a series of interactive exercises similar to the test items, and supplementary exercises. An illustrative example is presented in Fig. 2. Here, students can move a slider on a position-time graph, observe the tangent line at that point, visualize its slope, and see how the velocity-time graph is constructed as the point is moved on the graph. They can then use the graph to answer a set of multiple-choice questions (not shown). A parallel item in a purely mathematical context was provided just before.

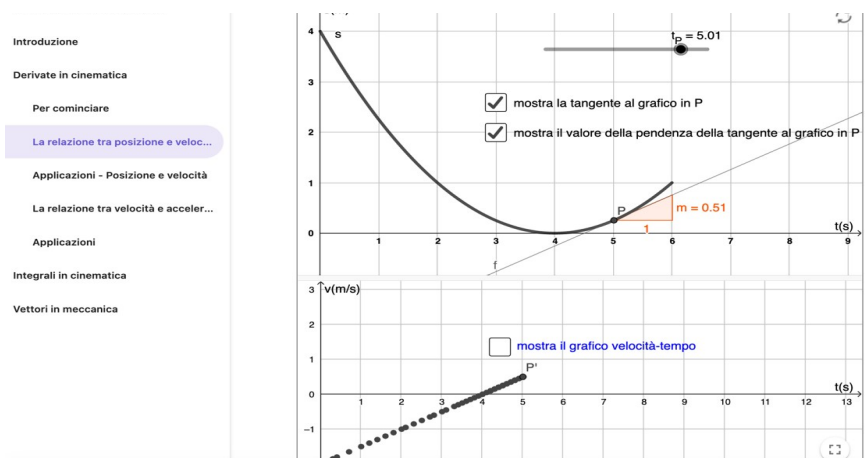

**Fig. 2**. An example from the interactive online learning modules developed from the TCV-MP test.

Since 2018, the test is made available for students every year at the beginning of their Physics 1 courses. For students, the test serves as a self-assessment tool, based on which they can identify the areas requiring more practice. The online modules are also provided as self-training tools. The results of the test are shared with the instructors of each course, so that they can gain insights into the situation of their class and possibly devise interventions. As an example of the information shared with the instructors, Fig. 3 shows the scores distribution for the degree courses that participated in 2022 and a single-item analysis for one of the degree courses (Mechanical Engineering).

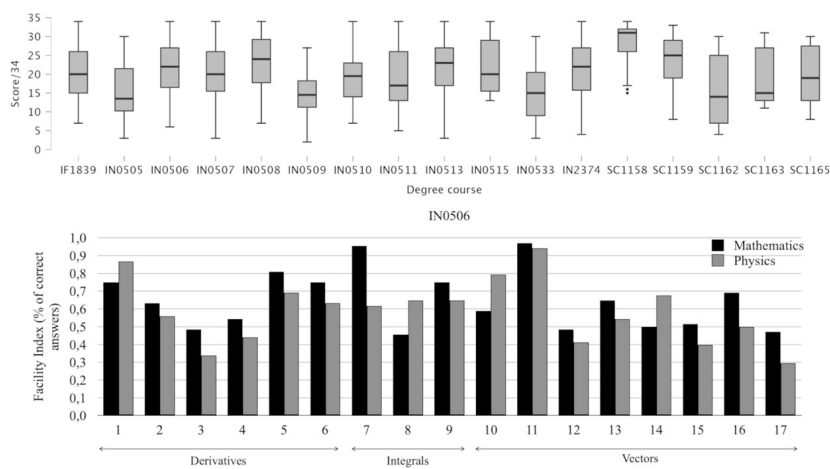

**Fig. 3.** Boxplots representing the scores distribution for the different degree courses that participated in the 2022 administration of the TCV-MP. Codes beginning with 'IN' identify courses belonging to the school of Engineering (e.g. IN0506 is Mechanical Engineering), while codes beginning with 'SC' identify courses belonging to the school of Science (e.g. SC1158 is Physics). 'IF' identifies an interfaculty course.

The TCV-MP has become an established tool offered to Science and Engineering students and instructors at our university. However, the number of participants has lowered since the first administration, stabilizing at approximately 700 students per year. The main factor affecting participation is the presence or absence of entire large-enrollment courses, which strongly depends on instructor engagement. While some colleagues consistently propose the test and encourage student participation in different ways, others are only superficially interested. The instructors are continuously changing and many of the new ones have not participated in the initial process. The involvement of instructors with different ideas about physic teaching and the role of physics education research remains a challenge.

In 2019, a modified version of the TCV-MP was developed for students in the last two years of secondary school (12th-13th grade), with the aim of reinforcing their mathematization skills before they enter university (Lipiello et al. 2022). While the section about vectors was maintained with minor modifications both in the mathematics and in the physics context, the section on derivatives and integrals in the mathematics context was adapted to suit the students' level. In particular, derivatives were conceptualized as the slope of the tangent line, while integrals were conceptualized as the area under the curve. Following test administration, a learning path was organized employing online modules similar to the ones described above, together with other strategies designed for the specific student population. The test was then administered again as a post-test to evaluate the efficacy of the learning path. The pilot version of the project was conducted in 2019/20 with a group of self-selected students and was reiterated in 2020/21 with a more representative group. Fig.4 shows the scores distribution for the pre- and post-tests. The results are promising and support the effectiveness of the learning modules in fostering students' mathematization skills.

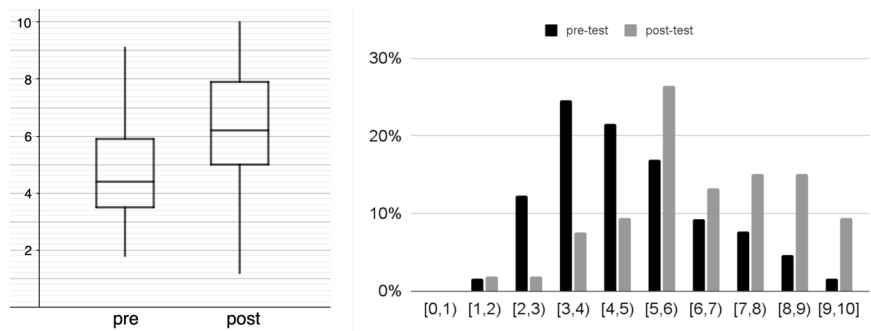

**Fig. 4**. Comparison between pre- and post-test results for the high-school version of the TCV-MP.

The test and the learning materials were used again in 2021 with 174 students from 7 classes (article in preparation). In 2022, a project was launched to extend the experimentation beyond the pilot school. Currently, five schools (including the original one) are experimenting the test and the learning path with different classes.

### 4.4 Conclusions

The project described here is part of our effort to improve students' mathematization skills in physics. We believe that the TCV-MP can be a valuable resource for both students and instructors. Combining the test with the online learning modules can further help students develop the mathematization skills they need to successfully engage in physics learning.

## 5. University Students' reasoning for Understanding of Moment of Inertia

### 5.1 Introduction

Undergraduate physics courses typically begin with mechanics, which is a review of high school physics for most of the students and one may expect that students will not have significant difficulties with it. However, as it has been shown, understanding the kinematics and dynamics of circular motion present difficulties for high school and *university* students (Klammer 1998). Therefore, students' difficulties grow up when a particle is substituted by a rigid body.

The research on the students' understanding of rigid body rotation at the University level has shown many difficulties. Frequently, students mix the concept of torque with the concept of force. In the same way, they do not establish any relation between the torque and the angular acceleration, they think that a constant torque results on a constant angular velocity and they show difficulties when taking into account the line of action of a force besides its point of application (Duman et al. 2015).

There is little research on rotational dynamics of a rigid body and its moment of inertia. This investigation shows that students often have difficulties in correctly relating the magnitude moment of inertia to other quantities involved in rotational dynamics such as mass or angular acceleration (Rimoldini & Singh 2005). However, we have found no research that deals with the understanding of the qualitative meaning of the moment of inertia in relation to the rotational motion of a body, nor with the relationship between the use of the analytical definition of the moment of inertia and the meaning with which it is used. Therefore, in the presented investigation we analyse the understanding of students in introductory physics courses of the qualitative meaning of the moment of inertia as the "resistance" of the body in changing its state of rotation and its relation to the use of the operational definition in familiar situations in academic contexts.   -

*5.2. Experimental design and methodology*

We conducted a study with students from first course of engineering of the University of Basque Country (UPV/EHU). All students take two physics courses involving topics on mechanics electromagnetism during postcompulsory education (16–18 years old). The students were randomly distributed among the first-year engineering groups. We gave students an open-ended questionnaire (7 questions) after they had studied the rigid body rotation, where rotation was considered to be around a fixed axis in space.

In order to validate the questionnaire, once it was prepared, three professors concluded that the objectives of each question were the ones we aimed for and were clear. Moreover, we carried out a draft test with 25 first-year course students, which confirmed that students had no problem understanding how the questions were formulated. Finally, the questions were included in the first-year students' test in the form of a post-test for first-year students who had already completed the semesters of Mechanic. Students' responses were analysed and classified in different explanatory categories that emerge on the analysis. The categorization process was validated by three researchers, and the classification of students' answers into categories was supported by Cohen's kappa ($\kappa$ = 0.92). The questionnaire was given to 97 first-year university students.

The questions on the questionnaire are familiar to students in the academic context and are usually mentioned in textbooks as examples of rigid body rotation. In Figure 1 we present one of questions related with the qualitative comprehension of the moment of inertia as the resistance to change the rotational motion of a body. This question will serve as an example to show how we carried out our investigation on students' understanding of the moment of inertia.

**Question Q1**. Let us suppose a situation where two balls with the same mass and the same radius joined by a rod of negligible mass can rotate with respect to an axis located in the middle of the rod. The following three cases are presented:

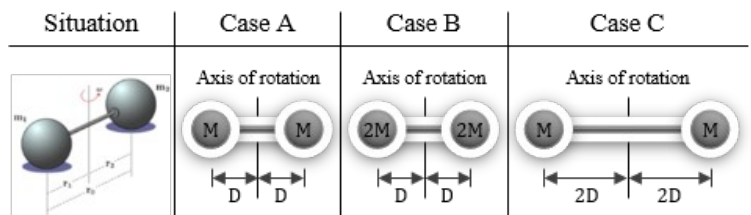

In which case will it be easier to rotate the system? Sort the cases from greatest to least according to how easy it is to rotate the system. EASIEST ____ ____ ____ HARDEST. **Argue your answer**.

**Figure 1.** To inquire about moment of inertia, a rotational motion of two balls joined by a rod is proposed to students. They need to analyse the moment of inertia of the system in three different cases and sort those cases according to easiness to rotate with respect to the axis. The concept moment of inertia is not mentioned in the question.

Question Q1 aims to investigate whether students understand the meaning of the moment of inertia in a rotating particle system. According to the theoretical framework of classical physics, the moment of inertia measures the resistance that a body that rotates around an axis opposes to the change of rotational motion and is defined as: $I = \sum m_i r_i^2$. The moment of inertia about an axis has an analogous role to that of mass in translational motion.

The three different cases in the question Q1 differ the distance from the balls to the axis of rotation and the balls' mass, so that the students can make an analysis when justifying their choice of ordering the cases according to the analysis of variables that influence on the change of rotation, values of moment of inertia and its meaning.

*5.3. Results*
We first present the five categories that emerged from the analysis of the answers to question Q1 regarding the concept of moment of inertia (see Table 1). Then, we show a summary of the results, highlighting the main findings and discuss their implications.

Category A includes explanations that imply an understanding of moment of inertia as a resistance of a body to change its rotation state. We separated this category into two subcategories A1 and A2. Most of the answers in category A take into account both variables that influence the moment of inertia explicitly, the mass and the distance to the rotation axis, in a proper way according to the definition of moment of inertia $I = \sum m_i r_i^2$ (subcategory A1). The other responses in category A make an explicit qualitative analysis of the variables without putting any equation in, they only indicate the calculation of the equation to justify their answer or they do not get the calculation right due to some small mistake (subcategory A2).

**Table 1.** Categorization of question Q1. Name and description of each category are shown in the first and second columns from the left respectively. The column on the right show the percentages of the answers classified in each category. (N=97).

| Cat. | Explanation | % of answers |
|---|---|---|
| A1 | They understand that the moment of inertia is related to the difficulty/ease of rotating a system and calculate it correctly. | 34 |
| A2 | They understand that the moment of inertia is related to the difficulty/ease of rotating a system but they do not compute it correctly. | 11 |
| B | Arguments based on the idea that the difficulty of rotation depends on the mass and/or the distance to the axis. | 22 |
| C | Affirmations based on rote resources. | 24 |
| D | Incoherent. | 4 |
| E | No answer. | 4 |

Category B include responses that correctly calculate the value of the moment of inertia but do not analyse the two variables, mass and distance from the axis of rotation, together. They do a functional reduction giving priority to one of the variables, without justifying why is one of them prioritized over the other one. This could be related with linear causal explanations that establish a chain of simple causes in which any change is the result of a previous change, and in turn, is the cause of the subsequent effect, For example:

"*The axis of rotation will be on the centre and the balls on the edges, the closer the balls from the axis the easier will be to turn the system.*" (Student No. 3)

Around 25% of the answers (see Table 1) were explanations that focus their analysis on one of the two variables (mass and distance to the axis of rotation) that influence the moment of inertia and were classified as category C. The reductionist analysis of the moment of inertia, in this case, leads to incorrect answers. These incorrect answers reflect a lack of understanding of the concept of moment of inertia.

"*The closer the turning objet from the axis of rotation, the harder it will be for the system to turn because of its smaller moment of inertia. If the distance between the axis of rotation and the object is the same, the smaller the mass the greater its moment of inertia, then it is easier to turn the system.*" (Student No. 5)

*5.4. Discussion and Conclusions*

Our study includes individual student's questionnaire using open-ended questions for inquiring students' learning on momentum of inertia. Here we have presented one of the questions on the questionnaire, but the student difficulties we have described are common to other questions. We have described some common mistakes students make when answering questions about moment of inertia in a context of rotation of a particle system (Q1). A third of the students answer question Q1 correctly (category A1), and we understand that about half of the students show a correct understanding of the moment of inertia and its meaning (categories A1 and A2). However, we note that many others have difficulties in analyzing the influence that moment of inertia has on a body's rotation. Many students do not identify the moment of inertia as a quantity that measures the resistance of the system to the change of rotation, and then it is likely that they have difficulties in establishing relationships between moment of inertia of a body and its angular acceleration of rotation. It occurred for around a third of students in Q1 (category C). Furthermore, some students, although a minority of cases, correctly indicate the value of the moment of inertia, the lack understanding of its meaning leads to erroneous reasoning patterns. We find valuable the idea of moment of inertia as the resistance of the system to the change of rotation, because the learning difficulties with the idea of "inertia" also extend to similar ones with linear motion in high school students (Lehavi & Galili 2009).

The study also shows that there can be a learning progression about calculating and understanding the moment of inertia in comparison with high school students (24% for Q1) (Olazabal et al. 2021). Beginning with students who are not able to calculate the moment of inertia, nor understand its meaning (categories C), going through other students who know how to calculate it from the analysis of the mass and distance to the axis of rotation or either qualitatively understand the meaning of moment of inertia (category B). Finally, those who understand the meaning of the moment of inertia and are able to calculate it and understand it properly as a resistance to change body's state of rotation (category A).

In traditional instruction, it is first taught to calculate the moment of inertia separately as a mathematical integration problem and later, the moment of inertia is analyzed in the context of Newton's second law for rotation. We show that a significant part of students does not relate the moment of inertia with the analysis of change in rotational motion of a body. This suggest that instruction could be more effective if the moment of inertia is taught from the beginning in relation to its meaning of opposing the change of rotation of the system and, being aware the problems that the students will have in the study of the second Newton's law of rotation.